\documentclass[conference]{IEEEtran}
\IEEEoverridecommandlockouts
\usepackage[english]{babel}
\usepackage{graphics, graphicx}
\usepackage{amsmath, amsthm, amssymb, amsfonts}
\usepackage{algorithm}
\usepackage{algpseudocode}
\usepackage{multirow}
\usepackage{url}
\usepackage{soul}  
\usepackage{tabularx}
\usepackage{caption}
\usepackage{subcaption}
\usepackage{comment}
\usepackage{tikz}
\usepackage{cite}
\usepackage{xcolor}
\usepackage{tabu}
\usepackage{tabularx}
\usepackage[nocomma]{optidef}
\usepackage{soul}

\usepackage{enumitem}
\definecolor{gold}{rgb}{0.85,.66,0}

\newcommand{\bPh}{\boldsymbol{\Phi}}
\newcommand{\bth}{\boldsymbol{\theta}}

\DeclareMathOperator*{\argmax}{arg\,\max}

\def\BibTeX{{\rm B\kern-.05em{\sc i\kern-.025em b}\kern-.08em T\kern-.1667em\lower.7ex\hbox{E}\kern-.125emX}}

\title{LSTM-ACB-Based Random Access for Mixed Traffic IoT Networks}

\author{

\IEEEauthorblockN{{Herman Lucas dos Santos}}
\IEEEauthorblockA{\textit{Electrical Eng. Dept.}
\textit{UEL}\\
Londrina, Brazil.\\
hermanlds@gmail.com
\vspace{-4mm}}

\and

\IEEEauthorblockN{{Jo\~ao H. Inacio de Souza}}
\IEEEauthorblockA{\textit{Electrical Eng. Dept.}
\textit{UEL}\\
Londrina, Brazil.\\
joaohis@outlook.com
\vspace{-4mm}}

\and

\IEEEauthorblockN{Jos\'e Carlos Marinello Filho}
\IEEEauthorblockA{\textit{Electrical Eng. Dept.}
\textit{UTFPR}\\
Corn\'elio Proc\'opio, Brazil.\\
jcmarinello@utfpr.edu.br\vspace{-4mm} }

\and

\IEEEauthorblockN{Taufik Abr\~ao}
\IEEEauthorblockA{\textit{Electrical Eng. Dept.}
\textit{UEL}\\
Londrina, Brazil.\\
taufik@uel.br
\vspace{-4mm}}

\thanks{This work was supported by the National Council for Scientific and Technological Development (CNPq) of Brazil under Grants 405301/2021-9, 141485/2020-5, and 310681/2019-7.}
}

\begin{document}

\maketitle

\begin{abstract}
We propose a novel random access (RA) protocol that accounts for the network traffic in mixed URLLC-mMTC scenarios. By considering an IoT environment under high mMTC traffic demand, we model the traffic of each service using realistic statistical models, with the mMTC and URLLC use modes presenting a long-term traffic regularity. A long-short term memory (LSTM) neural network (NN) is used as a network traffic predictor,  enabling a traffic-aware resource slicing (RS) scheme, 
aided by a contention access control barring (ACB)-based procedure. 
The proposed method combines a grant-based RA scheme, where it is introduced an intermediate step in grant-free RA, to deal with collisions. The protocol presents a small overhead, supporting a higher number of packets in a frame thanks to the congestion alleviation enabled by the ACB procedure. Numerical results show the effectiveness in combining the three procedures in terms of accuracy for traffic prediction, resource utilization and channel loading for RS, and increased throughput.
The comparison with a grant-free benchmark reveals substantial improvement in system performance.  
\end{abstract}


\section{Introduction}
Mixed network traffic involving the URLLC and mMTC services is challenging, since the former needs upon-requisition access, while the latter produces huge amounts of sporadic packet transmissions \cite{Petar2018,Zhang2021,Alsenwi2021,
Anand2020,Di2019}. Designing networks that support the co-existence of services demands studies on the traffic pattern and the development of strategies to allocate the resources aiming to deliver the expected performance levels. A comprehensive evaluation of the impact on the co-existence of use modes under orthogonal multiple access (OMA) against non-orthogonal multiple access (NOMA) is discussed in \cite{Petar2018}.  Statistical models for the network traffic in scenarios with the co-existence of services can be found in \cite{Alsenwi2021,Zhang2021,Anand2020,Thota2019}. Given the heterogeneous network traffic and the different \textit{quality of service} (QoS) required by each use mode, intelligent resource allocation schemes must be developed.

To fulfill QoS for different use modes in the same system requires adapting the available resources for properly supporting 
the users, namely resource slicing. RS consists in separating the transmission resources, in general, \textit{time} (T) and \textit{frequency} (F), to serve different services. Intuitively, URLLC demands high bandwidth to ensure fast and reliable transmissions, while mMTC requires lots of small resource blocks (RBs) to ensure high connectivity.
To tackle these heterogeneous demands, 3GPP release 15 introduced the concept of {\it flexible numerology}, enabling RS schemes by allowing sub-carrier spacing larger than the traditional 15 kHz of \textit{long-term evolution} (LTE) to grant more bandwidth for transmissions \cite{Memisoglu2021}. 
Different RS strategies are proposed in \cite{Tun2020,Alsenwi2021} to efficiently allocate the T-F resources in scenarios with mixed traffic.

{\it Access control optimization} (ACO) by exploiting the estimation of the network \textit{backlog}, \textit{i.e.}, the number of users trying to access the network at each frame, can be employed to attain simultaneously an improved traffic control and resource utilization 
\cite{Sim2020,Bui2020}. ACO procedures can be classified in: a) machine learning (ML)-based tools; and b) classical non-ML-based techniques. One efficient ACO procedure is the ACB scheme, which implements a broadcast message used by the users to decide whether or not to access the network. 
\cite{Sim2020} proposes a deterministic barring factor for networks where the users have different access priorities. On the other hand, \cite{Di2019} develops an ACO-ML method, which can manage the traffic with the help of a traffic predictor, enabling online adaptation and better precision at the cost of a higher 
complexity.

This paper proposes a hybrid RA protocol assisted by an LSTM traffic predictor and a contention ACB-based scheme for mixed mMTC-URLLC 5G use modes. {The proposed method combines three structures: {\it i}) a traffic predictor that can predict the arrivals of both traffic types separately;  {\it ii}) a T-F RS scheme supporting both use modes under specific packet size and latency restrictions; and {\it iii}) a novel three-step protocol, a modification of the classical grant-free structure with a collision management procedure that can alleviate network congestion. 
We propose a heuristic to manage and assign the base station (BS) resources aided by an LSTM-based traffic predictor, and relying on alternative use of ACB with estimation of the number of colliding users at each channel.  

\vspace{-2mm}
\section{System Model}
\label{sec:SystemModel}

Consider a time-division duplex (TDD) M-MIMO communication system with different types of service requirements, constituting a mixed URLLC-mMTC mode traffic. The environment is composed by a cell containing a centralized BS serving $\mathcal{K} = \{K^{\text{m}},K^{\text{u}} \}$ UEs, where $K^{\text{m}}$ is the number of UEs in mMTC use mode and $K^{\text{u}}$  is the number of UEs operating under URLLC mode; the former class 
{ask for connecting} very sporadically {(very low activation probability)}, but the number of mMTC devices is huge. 
\vspace{1mm}
\noindent\textbf{\textit{System Traffic}}: We adopt mMTC traffic as the major part of the users in the system with packet size of $P^m$ bytes, while the rest of UEs operates under URLLC mode with packet size of $P^u$ bytes, and assuming $P^m > P^u$ \cite{Alsenwi2021}. mMTC devices require activation as an uniformly distributed variable with activation probability $p$, also with a set of $K_m^p$ devices generating one packet every $T_m$ frames, characterizing both random and periodic traffic types \cite{Jiang2019}; {the URLLC traffic} as a repeating Beta distribution in a period of $T_u$ frames \cite{Liu2022}.  The number of mMTC and URLLC devices arrivals in the $t$-th T-frame is defined respectively as:
\vspace{-2mm} 
\begin{multline}
{\dot{K}^m} =  \sum\limits_{i = 1}^{K^{\text{m}}} \left[\mathcal{U}[0,1) \geq (1-p)\right]_i + K_m^p \sum\limits_{k=-\infty}^\infty \delta (t-kT_p)
\label{eq:mMTCTrafficModel} 
\end{multline}
\vspace{-4mm}
\begin{multline}
{\dot{K}^u} = \sum\limits_{i = 1}^{K^{\text{u}}} \left[\mathcal{U}[0,1) \geq 1- \frac{t^{\alpha-1}(T-t)^{\beta-1}}{T^{\alpha+\beta-1}\text{Beta}(\alpha,\beta)}\right]_i
\label{eq:URLLCTrafficModel}
\end{multline}
\noindent where $\delta$ is the impulse function, $\text{Beta}(\alpha,\beta)$ is the Beta function, and $\alpha$ and $\beta$ are the parameters of the curve. Besides, we define the number of arrivals summed related to the re-transmissions from previous frames, \textit{i.e.} the \textit{network backlog}, as $\breve{K}^{\text{m}}$ and $\breve{K}^{\text{u}}$ for mMTC and URLLC, respectively.  Each UE has distinct \textit{uplink} (UL) transmit power to compensate the pathloss differences.

\vspace{1mm}
\noindent\textbf{\textit{Resources Slicing in 5G New Radio}} (NR)
introduced the numerology in {\it sub-carrier spacing} (SCS) $\Delta f = 2^\mu \times 15$ kHz, with $\mu = \{0, 1, \dots, 4\}$ the numerology factor\footnote{$\mu>2$ is only applicable in millimeter-wave frequencies} \cite{Liu2022}, which is a useful feature for reducing {\it transmission time interval} (TTI), once more bandwidth is available, more information can be carried under the same time interval. A frame in 5G NR has a duration of 10 ms divided in $S$ time-slots. $S$ can assume values from $10$ to $40$, when $\mu=2$. The numerology can be used to either reduce the time slot duration or to increase the bandwidth, carrying more information. A typical TTI carries $\nu = 14$ OFDM symbols per RB {(single time-slot and sub-channel)}. Then, the expressions for TTI and, analogously, the \# symbols under a 1 ms time slot can be derived as \cite{Liu2022}:
\vspace{-4mm}

\begin{equation}
\text{TTI} = \frac{{N}_{\text{sym}}}{2^\mu \cdot \nu }\,\, \text{[ms]},\quad \text{or} \quad N_{\text{sym}} =  2^\mu \nu \,\,\,\,\text{[symbols/ms]}
\label{eq:TTI}
\end{equation}
\vspace{-4mm}

Each RB consisting in 1 ms, 15 kHz SCS time-frequency {(T-F)} block. Hence, one RB has 180 kHz of available bandwidth. A frame $t$ consists in $F \times S$ RBs, subject to the available bandwidth 
The BS can apply the numerology to create channels with higher transmission rate, \textit{i.e.} lower latency, creating sub-channels and assigning them to dedicated channels, {\it e.g.}, Fig. \ref{fig:Numerology}.  Let us consider a fixed minimum channel size of ten $1s\times 1f$ RBs to ensure the packet size transmission. Besides, the URLLC channels 1 and 2 are composed by three $\mu=2$ sub-channels each, mMTC channel 1 use $\mu = 0$ (minimum SCS usage), mMTC channels 2 to 4 use $\mu = 1$ (typical). The 6th to 8th mMTC channels use $\mu = 2$ (granting faster transmission, but less efficient in terms of resource allocation). In Fig. \ref{fig:Numerology}, the colors refer to the numerology or subcarrier size.  

\begin{figure}[!htbp]
\vspace{-4mm}
\centering
\includegraphics[width=.8\linewidth]{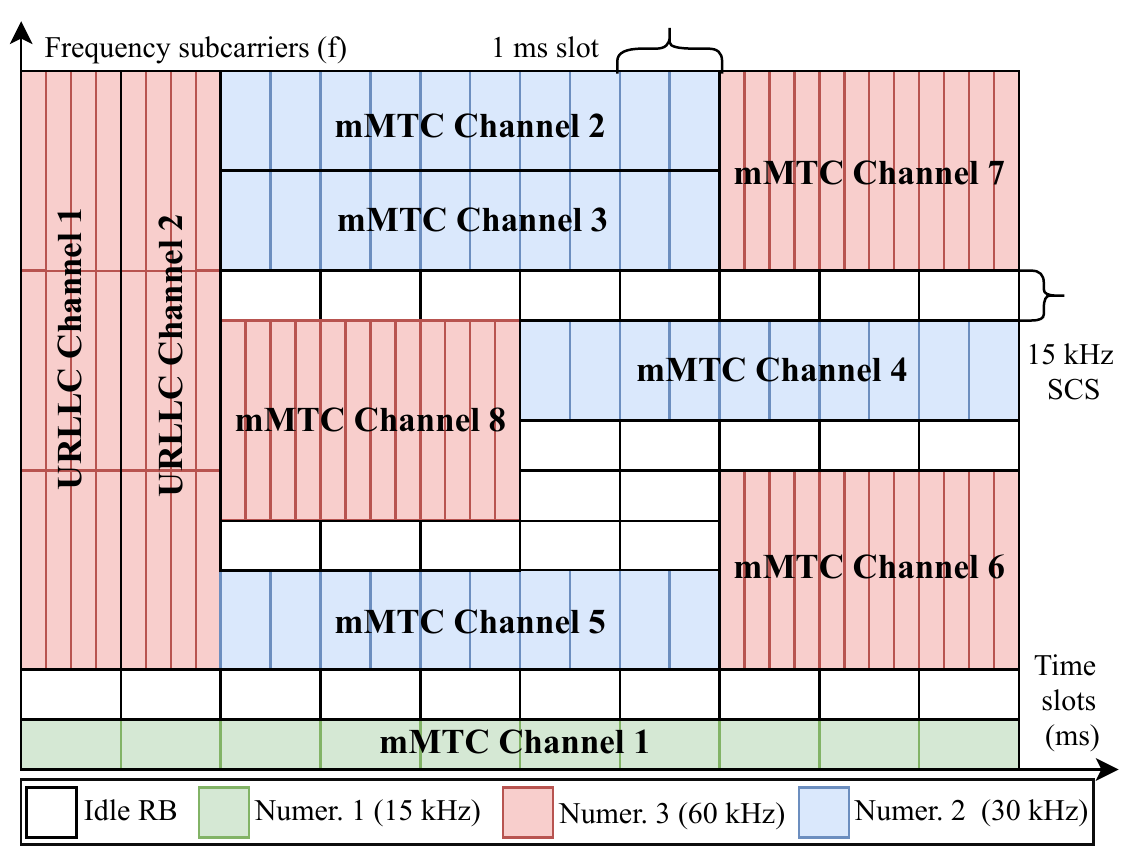}
\caption{\small Example of {T-F} RS to suit mMTC-URLLC mixed traffic under different numerology usage.
}
\label{fig:Numerology}
\end{figure}

\vspace{-3mm}
A set of {T-F} resources assigned to a specific user represents the $l$th channel, which is composed by one or more sub-channels, represented by the $\bPh_l$ matrix, in which each of them can accommodate one user transmission, with $\bPh_l^u$ representing the $l$th dedicated URLLC channel and $\bPh_l^m$ the $l$th mMTC one, while $\bPh_0$ represents the matrix containing the RBs. Notice that different modes do not compete for the same resources once the channels are dedicated (URLLC or mMTC modes). The {$F\times S$ dimensional {\it assignment}} matrix $\bPh$ contains $\phi_{s,f}=\{0,1\}$ elements indicating which {time-frequency slots  RBs} are assigned to the $l$th channel. Based on the actual backlog, the BS tries to accommodate all the $\breve{K}^{\text{m}}+\breve{K}^{\text{u}}$ users in orthogonal channels. The set of channels {represents the union of} two sub-sets, $\mathcal{L}^u$ and $\mathcal{L}^m${; hence,} $\mathcal{L} = \{\bPh_0,\, \bPh_1^u, \bPh_2^u, \dots, \bPh_{L^{u}}^u, \bPh_1^m, \bPh_2^m, \dots, \bPh_{L^{m}}^m\}$,
where the cardinality $|\mathcal{L}|=L=L^u+L^m$.
The frequency resource is represented as a column vector $\boldsymbol{\phi}_f=[\phi_{1,j}\, \phi_{2,j}\,\ldots \phi_{F,j}]^T$, {and}  $\boldsymbol{\phi}_s = [\phi_{j,1}\, \phi_{j,2}\,\ldots \phi_{j,S}]$ represents a time-slot row vector.


\section{Sub-Problem  Formulation and Solution}\label{sec:Problem}
\subsection{Backlog Prediction  {$(\mathcal{P}1)$}} \label{subsec:BacklogPred}
The backlog structure is represented 
{by} $N^t = \breve{K}^{\text{m}} + \breve{K}^{\text{u}}$ users that becomes active in the $t$th frame, either as a new arrival or repetition due to previous failed attempt. 
Each channel {can be defined by} 
three {states:} {\it i}) \textit{Success}, when a UE selects a channel solely, \textit{i.e.} a collision did not occur; {\it ii}) \textit{Collision}, when two or more devices select the same channel simultaneously; or 
{\it iii}) \textit{Idle}, when no device selected the channel. 
The  T-F RB channels in the same frame $t$ are represented in Fig.  \ref{fig:Numerology}. Channels are reserved for each class and the BS can maintain a record of the states at each frame:
\begin{equation}
 O^{t} 
=  \big\{\{V_s^{u,t},V_c^{u,t},V_i^{u,t}\},\{ V_s^{m,t},V_c^{m,t},V_i^{m,t}\}\big\}
\end{equation}
{referring to the number of access attempts of users mMTC ($m$) and URLLC ($u$),} for success ($s$), collision ($c$) and idle ($i$) dedicated channels in the $t$th frame. The set of previous observations, from the $(t-T_\textsc{w})$th to $(t-1)$th frame, where $T_\textsc{w}$ is the memory size, is named \textit{history} at $t$th frame:
\begin{equation}
\mathcal{H}^t = \{O^{t-{T_i}}, O^{t-(T_i-1)}, \dots, O^{t-1} \}
\end{equation}

The {\it number of active users of each class} ($n^u$ and $n^m$), as well as the {\it total backlog} in a specific time-frame can be estimated as the conditional probability \cite{Jiang2019} by solving:
\vspace{-6mm} 

\begin{eqnarray}
\hat{K}^{u,{(t)}}& = \underset{n^u \in \{0,1,\dots,K^u\}}{\argmax}  \mathbb{P} \{N^{u,{(t)}} = n^u|\mathcal{H}^{u,(t-1)}\} \nonumber \\
\hat{K}^{m,{(t)}} & = \underset{n^m \in \{0,1,\dots,K^m\}}{\argmax}  \mathbb{P} \{N^{m,(t)} = n^m|\mathcal{H}^{m,(t-1)}\} \nonumber\\
   \text{s.t.} & \hat{K}^{(t)}= \hat{K}^{u,(t)}+\hat{K}^{m,(t)} \label{eq:backlogPred}
\end{eqnarray}
\vspace{-6mm}

\noindent One task consists in predicting the number of users {per use mode} that will require access in the next T-F frame  $t$, {\it i.e.,} $\hat{K}^{u,(t)}$ and $\hat{K}^{m,(t)}$.  The BS has information {on} the number of active UEs.
Hence, the prediction can be employed solving the conditional probability problem in Eq. {\eqref{eq:backlogPred}}, by finding the optimum number of URRLC and mMTC users, term $n$ per class, with maximum probability. This is a supervised data-driven problem since it relies in training, \textit{i.e.} the probabilities are derived using observations and desired outputs. 

\vspace{1mm}
\noindent\textbf{\textit{LSTM-aided Traffic Predictor Solution}} $(\mathcal{S}1)$. 
Traffic network estimation can be modeled as a Bayesian probability inference problem, 
once it follows a distribution and the statistics can be captured from previous observation.  As the traffic configuration becomes more complex, ML techniques become suitable because of the high processing power, and it can be employed to capture several previous frames with low-complexity \cite{Jiang2019}. In hybrid URLLC-mMTC traffic scenario, the BS does not know how many users want to become active at each time frame, the only available information are the three detectable states of each channel, and, in occurrence of collisions, it is not known how many users have collided, once the BS is not able to decode the message in such case. 
A time series forecaster is composed by recurrent NNs running {\it long short-term memory} (LSTM), once it takes advantage of a set of previous observations, assuming that the traffic has periodic characteristics. 
The deployed LSTM is composed by several layers containing a forget gate, an input gate, and an output gate, where the latter is connected to the subsequent LSTM layer, and a softmax output function. 
The LSTM layers receive the previous observations $[\mathcal{O}^t, \mathcal{O}^{t-1}, \dots, \mathcal{O}^{t-T_0}]$, where $T_0$ is the number of layers in the NN, {and the softmax function outputs  the backlog traffic prediction, in terms of} $\mathbb{P}\{\hat{N}^{t+1} = n|{\mathcal{H}}^t,{\bth}\}$, {\it i.e.} the conditional probability of $n$ activations occur given previous $T_0$ observations under a configuration parameters {$\bth$}, {which are} optimized at each iteration till convergence.


\vspace{-1mm}
\subsection{Resource Slicing {$(\mathcal{P}2)$}}
The resource allocation must maximize the number of served users with orthogonal resources only,  satisfying the URLLC UEs TTI, and the different packet size for URLLC and mMTC users. Also, 
the sub-channels should be contiguous in T-F. Formally, the binary assignment problem is defined in eq. \eqref{eq:RBConfig},  where $\omega^u$, $\omega^m, {\omega^p} \in[0; \, 1]$, with $\omega^u >\omega^m \geq \omega^p$, and $\omega^u + \omega^m + \omega^p=1$ are priority scaling factors to URLLC and mMTC received packets and penalty term for not fulfilling the demand for channels, respectively; {$Z = \frac{F\cdot S - \iota^u\cdot K^u}{\iota^m}$, where} $\iota =  \left(\frac{P^i}{\log_2M^i}+\xi \right)\cdot \frac{1}{\nu}$ is the package size in symbols for $i \in {u,m}$, $\boldsymbol{\rho}^u =[\rho^u_1,\dots, \rho^u_{L^u}] \text{ and } \boldsymbol{\rho}^m = [\rho^m_1, \dots, \rho^m_{L^m}]$ are the {received} power gain vectors associated to the URLLC  and mMTC channels, respectively,  {$\mathbf{e}_1=\mathbf{1}^{1\times F},\, \mathbf{e}_2 = \mathbf{1}^{S\times 1}$}, and $\odot$ stands for the Hadamard product; $\nu$ is the number of symbols per RB, $\xi$ is total protocol overhead in symbols, $P$ is the packet size associated to each mode in bits, and $M$ is the modulation order. %
The {$\mathcal{P}2$} solution must optimize the {RBs usage, assuming known the CSI of all T-F sub-channels, given the priority factors $\omega^u$, $\omega^m$,} and subject to the penalty if the number of assigned channels is lower than the number of users under the T-F grid limitation, represented by the variable\footnote{The variable $Z$ stand for the maximum possibility of channels given $L^u = \hat{K}^u$ and the T-F grid size.} $Z$, and given the constraints \eqref{eq:RBConfigc1}--\eqref{eq:RBConfigc7}; the {elements of matrix $\bPh$ } in  constraint \eqref{eq:RBConfigc1} must be binary; in \eqref{eq:RBConfigc2}, each URLLC channel occupies only one time RB to guarantee the delay requirement and prevent resource wastage; \textit{\eqref{eq:RBConfigc3}} and \textit{\eqref{eq:RBConfigc4}) the RBs in each sub-channel must spam in continuous time and/or frequency RBs} to ensure the transmission is not fragmented neither in T nor in F; \eqref{eq:RBConfigc5} \textit{the number of sub-bands must satisfy the $2^\mu$ numerology factor}; \textit{\eqref{eq:RBConfigc6} the assignment matrices should not overlap each other} (orthogonal resources); 
in \eqref{eq:RBConfigc7}, the sum of assigned RBs must satisfy the packet size requirement.

\vspace{-5mm}
\begin{maxi!}|s|[2]{\bPh}{\sum\limits_{\ell=1}^{L^u} \omega^u \rho_{\ell}^u + \sum\limits_{\ell=1}^{L^m}\omega^m {\rho}_\ell^m {- \omega^p\left[\Breve{K}-\min\left(|L|,Z\right)\right]}} {\label{eq:RBConfig}}{}
\addConstraint{\phi_{f,s}}{\in \{0,1\}}{\label{eq:RBConfigc1}}
\addConstraint{\sum\limits_{i=1}^F \phi_{i,s}^u }{\leq 1}{\label{eq:RBConfigc2}}
\addConstraint{\Big\{\begin{matrix}
\phi_{k,s} = 1, \\
0,                      \end{matrix}}
{\begin{matrix}
\ i \leq k \leq j \\
 \text{ otherwise} \end{matrix}}{\label{eq:RBConfigc3}}
    \addConstraint{\Big\{\begin{matrix}
    \phi_{f,k} = 1, \\
 0,      \end{matrix}}{\begin{matrix}
 \ i \leq k \leq j \\
 \text{ otherwise} \end{matrix}}{\label{eq:RBConfigc4}}
\addConstraint{\mathbf{e}_1\boldsymbol{\phi}_{f}}{=k\cdot 2^\mu \ \forall \ \mu = \{0,1,2\}, \  k \in \mathbb{N}}{\label{eq:RBConfigc5}}
\addConstraint{\mathbf{e}_1\left[\bPh_i \odot \bPh_j\right]\mathbf{e}_2 = 0 }{ \ \forall \  i,j,\ i \neq j}{\label{eq:RBConfigc6}}
\addConstraint{\mathbf{e}_1\bPh^{\text{i}}\mathbf{e}_2}{\geq \iota^i, \  i \in \{u,m\}}{\label{eq:RBConfigc7}}
\end{maxi!}
\vspace{-5mm}

\noindent\textbf{\textit{RS based on Heuristic \textsc{MaxRect} $(\mathcal{S}2)$}}. The problem in \eqref{eq:RBConfig} can also be interpreted as a \textit{bin-packing} problem, once there is a limited space, \textit{i.e.}, the T-F grid, and the channels can be packed as boxes with different sizes and should be arranged in a manner that the maximum of \textit{boxes} fit into this space \cite{Jukka}.
We adopt an heuristic approach based on the {\it maximal rectangles} {{\it bottom-left} (\textsc{MaxRect})} \cite{Jukka}, 
Algorithm \ref{alg:maxrect}.

\subsection{Congestion Control:  ACB $(\mathcal{P}3)$}


The BS broadcasts one or multiple ACB factors $p_{\text{acb}} \in [ 0,1 ]$ in system information broadcast (SIB2) message. Devices that want to become active must perform an ACB check, generating {a} random number $q \sim \mathcal{U}[0,1)$ and  {comparing it to the ACB factor} $p_{\text{acb}}$; 
if $q\leq p_{\text{acb}}$ then the device can proceed to the RA process, else it must wait $[0.7+0.1\cdot\mathcal{U}[0,\,1)]T_{\text{acb}}$, where $T_{\text{acb}}$ is the {\it barring time}, till next access attempt \cite{Bui2020}. Also, {a} parameter $W$ is employed to determine the maximum number of  access attempts a device can perform before it is shut down. 

In the ACB procedure, 
at each frame, a set of $\Breve{K}$ active users selects one of the $L$ available channels\footnote{For convenience, in this section we suppress the $m$ and $u$ indexes.}.  The probability of a success transmission, \textit{i.e.} a channel is chosen solely by one device, is $\binom{\Breve{K}}{1}\frac{1}{L}\left(1-\frac{1}{L}\right)^{\Breve{K}-1}$. Expanding for L possible channels, the number of successful transmissions is \cite{Sim2020}: 
\vspace{-1mm}
\begin{equation}
V_s = L\binom{\Breve{K}}{1}\frac{1}{L}\left(1-\frac{1}{L} \right)^{\Breve{K}-1} = \Breve{K}\left(1-\frac{1}{L} \right)^{\Breve{K}-1}
\end{equation}
To maximize the number of decoded packets, the ACB factor $p_{\text{acb}}$ must be chosen carefully to ensure $\Breve{K}^i \cdot p_{\text{acb}} \leq L^i, \ i \in \{u,m\}$. As a result, the optimal $p_{\rm acb}$ is the factor that maximizes the number of successful transmissions $V_s$,  obtained as $\frac{dV_s}{d\breve{K}} = 0$ 
%
%
Thus, for $n$ devices trying to access a single channel in the T-F grid, the optimal ACB factor is \cite{Sim2020}:
\begin{equation}
    p_{\rm acb}^{*} = 1-\min\left(1, n^{-1} \right)
\end{equation}
In the above optimal ACB factor derivation, 
perfect power control is assumed.
Hence, the BS can estimate precisely the number of users attempting to access every channel.

\vspace{-1mm}
\begin{algorithm}[!htbp]
\small
\begin{algorithmic}[1]
\Require{$\hat{K}^u$, $\hat{K}^m$, $F$, $S$, $P^u$, $P^m$}
\Ensure{$\bPh$}
 \While{$l \leq \hat{K}^u$}
\State Find in subspace $\mathcal{R}$ the bottom-left vertex\;
\State Assign $\big( \frac{P^u}{\log_2 M^u} + \xi \big) \cdot \frac{1}{\nu} $ freq. in 1-time-slot to $l$th channel\;
\State $l = l+1$
\State $\mathcal{R} = \mathcal{R} - l$ \EndWhile
 \While{$[l \leq (\hat{K}^u + \hat{K}^m)] \text{ {\bf AND} } [|R| \geq \big( \frac{P^m}{\log_2 M^m} + \xi \big) \cdot \frac{1}{\nu}]$}
\State Find in subspace $\mathcal{R}$ the bottom left vertex\;
\State Set $\mu$ = 0 and shape the channel \textit{box}\;
\State Assign the selected resources to the $l$th channels\;
\While{$l$th channel do not fit the resource grid}
\State $\mu = \mu+1$; recalculate the box\;
\If{$\mu = 2$}
\State Set $\mu = 0$; search for next bottom-left vertex in $\mathcal{R}$\;
        \EndIf
    \State $l = l+1$
    \State $\mathcal{R} = \mathcal{R} - l$ \EndWhile
    \EndWhile
 \caption{\textbf{\textsc{maxrect}} - Maximal Rectangles Algorithm}
 \label{alg:maxrect}
\end{algorithmic}
\end{algorithm}
\vspace{-4mm}

\subsection{{Full-Solution: LSTM-ACB-based RA Scheme with Slicing}}
\label{sec:RAScheme}


The proposed LSTM-ACB-based RA scheme, Fig. \ref{fig:H-RL RAprot}, combines the grant-free fast transmission and an ACB-based control to alleviate collisions in congested hybrid URLLC-mMTC traffic modes. 
Based on prioritized access control, the method should grant resources preferable to {prioritized} URLLC users, allowing them to transmit under the same {T-F} frame and satisfying the 1 ms latency constraint, while re-scheduling mMTC users through the ACB parameter.  
%


\begin{enumerate}[label={S.\arabic*},left=0mm,start=0]
\item  {\bf(SIB2)} System Information Broadcast -- Based on a belief state containing $\hat{K}^u$ and $\hat{K}^m$, the BS solves RA allocation $\mathcal{P}2$ to define the optimal  \# channels for each service.
\item {\bf(Msg1)} Preamble selection -- Users with data in their queue receive SIB2 message and initiate the RA procedure by transmitting the preamble selection message in UL.
\item {\bf(Msg2)} Barring factor transmission -- BS observes triplet $\{V_s, V_c, V_i\}$, broadcast the barring factor $p_{\rm acb}$ to each user; the colliding UEs {generating} a random number and compares with $p_{\rm acb}$ to determine if they proceed to Msg3.
\item 
{\bf(Msg3)} UEs that continue the RA procedure{, \it i.e.}, UEs that chosen the channels solely and users that pass the ACB {rule}, repeat their preamble and transmit their data.
\end{enumerate}
%

 \begin{figure}[htbp!]
\vspace{-3mm}
\centering
\includegraphics[trim=1.8mm 0mm 28mm 0mm, clip,width=.8\linewidth]{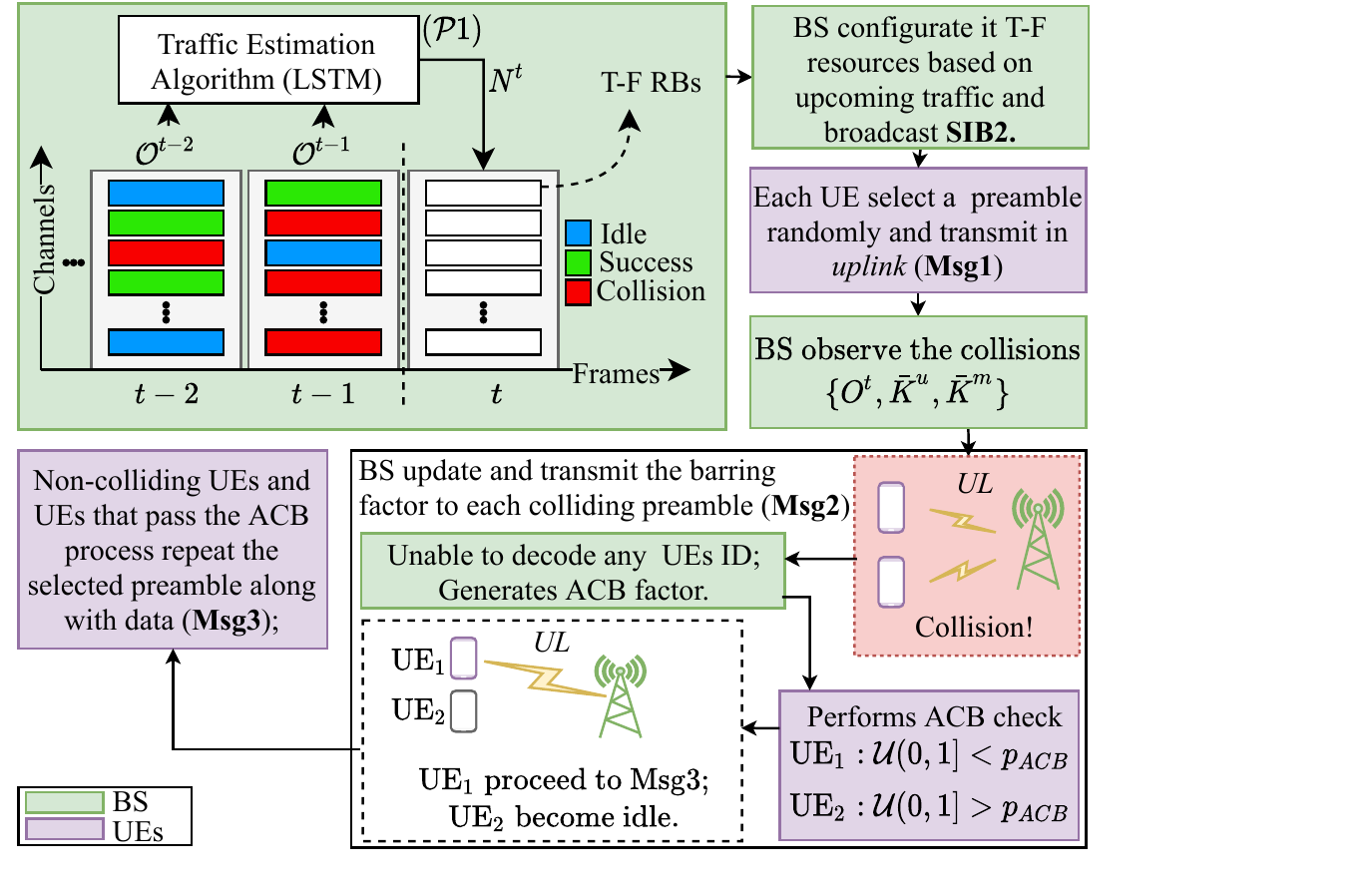}
\vspace{-3mm}
\caption{\small LSTM-ACB-based RA Scheme for hybrid traffic.} 
\label{fig:H-RL RAprot}
\vspace{-3mm}
\end{figure}

\section{Results and  Discussion}\label{sec:results}
\vspace{-3mm}

\begin{table}[!htbp]
\caption{Adopted values for system and channel parameters.}
\centering
\begin{tabular}{l|l}\hline
 \bf Parameter      & \bf Value  \\ \hline
URLLC UEs   & $K^u = 25$  \\
mMTC UEs    & $K^m = 1000$  \\
{\# Frames} & {$\mathcal{F}=1200$} \\ 
Periodic mMTC UEs & $K^p_m = 10$\\
    Periodic uRLLC frames  & $T_p = 10$    \\
   \# Time RBs per frame   & $S = 10$    \\
{\#} Frequency RBs per frame & $F = {50}$\\ 
System Total Bandwidth & ${9}$ MHz\\
URLLC packet size  & $P^u = 32$ bytes  \\
mMTC packet size   & $P^m = 200$ bytes \\
{ACB factor} & {$p_{\rm acb}\in\{0.2, 0.4, 0.6, \frac{1}{\bar{n}}\}$} \\
URLLC Mod. order       & $M^u = 4$     \\
mMTC Mod. order       &  $M^m = 256$   \\
\# OFDM symbols per RB        &$\nu = 14$    \\
Weights: URLLC, mMTC, Penalty & $\{\omega^u\, \omega^m\, \omega^p\}$ = {$\{.9\,\, .05\,\, .05 \}$}  \\
Protocol Overhead & $\xi = 5$ \\
\hline
{Monte-Carlo simulation (MCS)}& {$\mathcal{T}\in[10^2; \, 10^3]$ realizations}\\
\hline
\end{tabular}
\label{tab:SimulationParameters}
\end{table}

\vspace{-1mm}
\noindent{\bf Traffic Prediction}:  to evaluate the performance of the NN solely, we consider fixed channels and dedicate part of them to URLLC traffic, \textit{i.e.} the remaining (major part of them) are dedicated to mMTC traffic. We adopt 
a total of 54 channels and fix 5 {channels} for URLLC traffic. The LSTM is trained through 1200 epochs, with {$10
^4$} traffic samples and 20 layers. 
We consider the two triplets referring to which traffic type as the inputs for the LSTM, $\mathcal{H}^t$ and $T_w = 20$, \textit{i.e.}, the LSTM uses past twenty samples in each prediction, Section \ref{subsec:BacklogPred}. 
The traffic prediction via LSTM ML compared with the desired output via exact backlog reveals a mean-squared error (MSE) over $\mathcal{T}=10^3$  MCS  in the range of ${\textsc{mse}^m, \textsc{mse}^u}= \{10^{-6}, 10^{-5}\}$, for {{\it low} traffic} $K^m,K^u = \{500,13\}$ and ${\textsc{mse}^m, \textsc{mse}^u}= \{10^{-3}, 10^{-2}\}$, for {{\it high} traffic}  $\{K^m,K^u\} = \{2500,63\}$. Hence, the mixed {mMTC and URLLC} traffic prediction is feasible using the LSTM-NN.%


\noindent{\bf Resource Slicing} (RS) procedure $(\mathcal{S}2)$. 
 In the absence of RS, the BS must structure the channels to serve any user\footnote{{Numerology index = 2 (60 kHz SCS) and 16 RBs per channel.}}. In such a situation, the maximum number of channels is $\frac{F\cdot S}{\iota^m} = 31$ channels for $F=50, S=10, \iota^m = 16$, to be accommodated in the T-F grid. In practice, the upper bound is 30 due to limitations in channel formats. The heuristic \textsc{maxrect} disregarding the power gains is able to allocate more channels, especially when $\Breve{K}^u$ is larger, since the packet size $\iota ^u < \iota^m$ and the use of different channel formats for $L^m$ translates into better T-F resource use. This gain in number of channels directly impacts the  \# users served and
the backlog.

Fig. \ref{fig:ChannelLoading} depicts the {\it channel loading} (CL) for the use modes under the deployed traffic characteristic. CL is the ratio between the {\it active users} by the {\it number of available resources}: %
\begin{equation}
    \text{CL}^i = \frac{\Breve{K}^i}{|L^i|}, \,\, \ i \in {u,m}
\end{equation}
The quantity of channels the system is capable of allocating under the RS perspective is increased as more URLLC channels are employed, varying from 31 ($|L^u|=1$) to 41 ($|L^u|=40$).  Without RS, 
the maximum number of channels is 30, occupying 480 RBs, once every channel is a $\mu=2$ (60 kHz SCS) with 16 \textit{vertical} RBs each. This type of channel can accommodate either URLLC or mMTC users in terms of latency and packet size. Applying the RS scheme, the RBs usage is improved, taking advantage of smaller packets demand from URLLC and different numerology, and higher TTI for mMTC, granting the ability to assign more channels in the T-F grid.
Also, 
the CL performance using RS is capable of dedicating channels based on the upcoming traffic and optimize the resource utilization: 
line CL=$1$ in the subplots of Fig. \ref{fig:ChannelLoading} {\it i})  and {\it iii}). 
Moreover, the occurrence of collisions in the presence of RS and fixed T-F grid is shown in the subplots {\it ii)} and {\it iv)}. Besides, the {\it average} CL and {\it collisions} over 1200 frames are depicted in the right subplots of Fig. \ref{fig:ChannelLoading}. The \textsc{cl}=1 line for both use modes reveals suitable resources management, serving all users, which can be achieved with further coordination. {Such access performance was} obtained {assuming URLLC priority mode} for every $\{K^u,K^m\}\leq \{100,4000\}$.  Furthermore, the collisions is significantly reduced {under the proposed RS procedure; indeed,} the system still running, although congested, for a high loading ($K^m=1000$), as indicated in subplot {\it iv)}.
As the loading in the system increases, the congestion in priority URLLC use mode is mitigated.

\vspace{-.1mm}
\noindent{\bf ACB Congestion management} ($\mathcal{P}3$) is analysed under LSTM-ACB-based RA protocol (Msg 2 and Msg 3) and
 assuming a fixed $L=54$ channels (split into URLLC and mMTC use modes), varying the number of users, the number of reserved channels for each service, and applying the ACB-based contention resolution in Msg2.  
The performance is averaged over $\mathcal{T}=100$ MCS, and over $\mathcal{F}=10^3$ frames. The {\it normalized throughput} is adopted as evaluation metric:
\begin{equation}\label{eq:nThroughput}
\eta=\frac{V_s}{V_s+V_i+V_c}
\end{equation}
where $p_{\rm acb} \in \{0.2, 0.4, 0.6, 1, \frac{1}{\bar{n}^{-1}}\}$,  and $p_{\rm acb} = 1$ representing the system without the Msg2 and Msg 3, \textit{i.e.}, a {simple} grant-free protocol where every colliding user is shut down in current frame, while the \textit{optimal} ACB factor is calculated as $p_{\rm acb}^* = \bar{n}^{-1}$, by assuming that the BS perfectly knows how many users collided in each channel. Also, we vary the number of users from $10^3$ to $3\cdot10^4$, for mMTC mode, while URLLC follows a rule of proportionality. 
Also, we consider $T_{\rm {\textsc{acb}}} = 0$, \textit{i.e.} a colliding user will retry the RA procedure in next frame. We test the {\it channels reservation} as:\\ 
(a) $L^u = 8$, $L^m=46$, \,  {$\forall$\, subset} $\{K^m,K^u\} {= \{K^m,\frac{K^m}{{40}}\}}$;\\
(b) $L^u \in \{4, 5, \dots, 33, 34\}$ as $K$ increases.

\begin{figure}[htbp!]
\vspace{-3mm}
\centering
\includegraphics[width=1\linewidth]{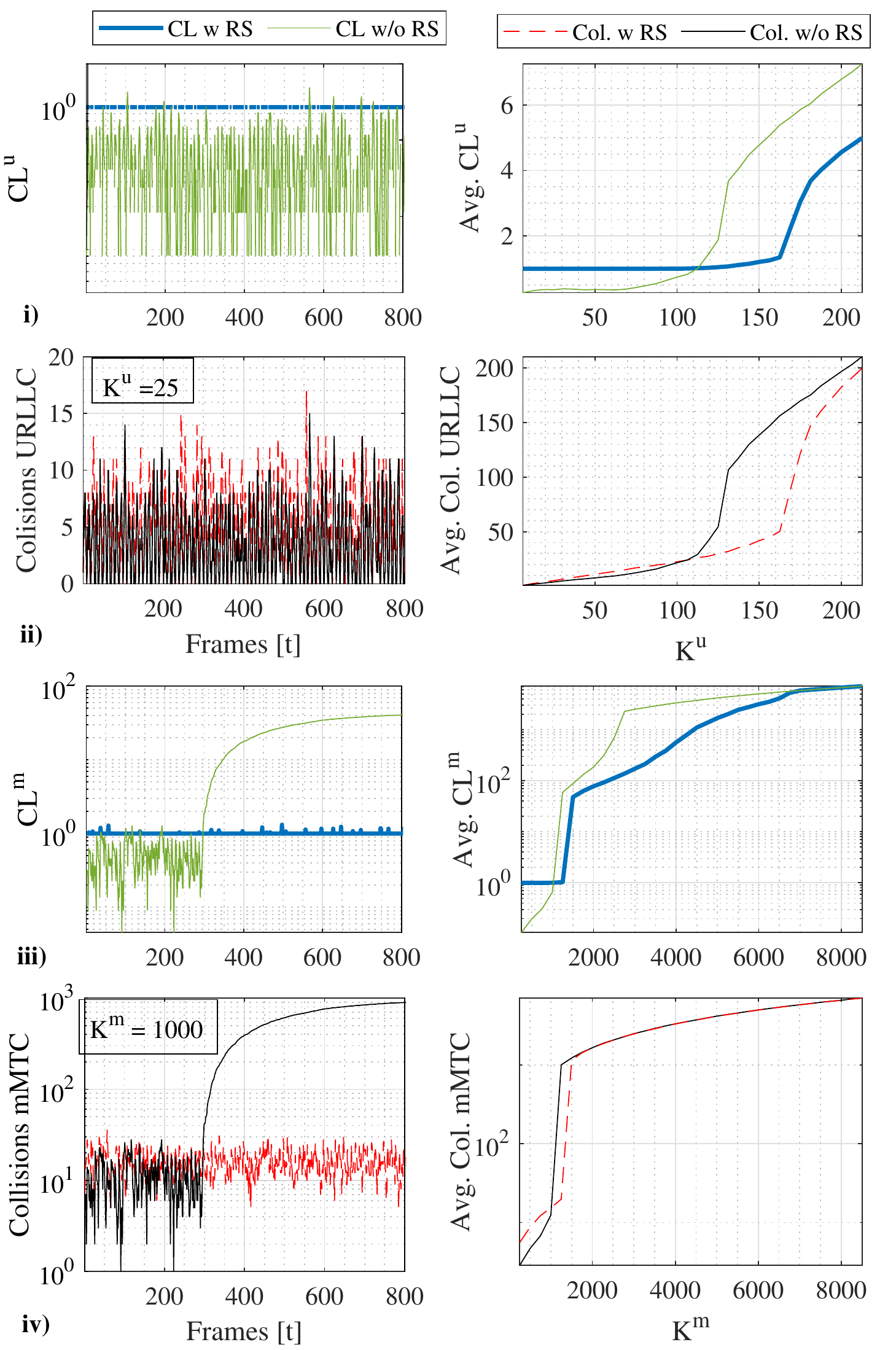}
\vspace{-5mm}
\caption{\small Channel loading performance  with RS (blue, red lines) and without RS (green, black lines) of channel loading and collisions for  URLLC in {\it i}) and {\it ii}), and mMTC in {\it iii}) and {\it iv}) for different \# users.}
\label{fig:ChannelLoading}
\vspace{-2mm}
\end{figure}

{Fig.} \ref{fig:HGP_Preliminary} shows the normalized throughput for increasing number of users in the given scenarios. {Hence, in the grant-free benchmark ($p=1$) operating under} $\{K^m = 4000, K^u = 100\}$, {or} higher $K$'s, the system {becomes} overload (high congestion), and none of the users from both {use} modes can access the network. It occurs due to {\it unsolved collisions} and {\it retransmission requests}, increasing the system backlog until there is more than one user attempting to access every channel in every frame. Applying the proposed protocol idea sketched in  Fig. \ref{fig:H-RL RAprot} (Msg2 and Msg3), the {\it congestion is alleviated}, and the performance increases significantly{, allowing the system be capable to allocate RBs for part of the UEs} in both use modes. 
Moreover, by adopting the optimal $p^*$, the performance is maintained for high number of users, say $K^m \gg 2000$. The channel reservation accordingly to the number of URLLC UEs {reveals} {remarkable improvements} in $\eta$ 
for every $p_{\rm acb}$. However, due to the traffic distribution throughout the frames, some of these channels are not utilized, specially in the start and final of the distribution periodicity 
($T_p =10$). 

\vspace{-2mm}
\begin{figure}[!htbp]
\centering
\begin{subfigure}[!htbp]{0.49\textwidth}
\centering
\includegraphics[width=\textwidth]{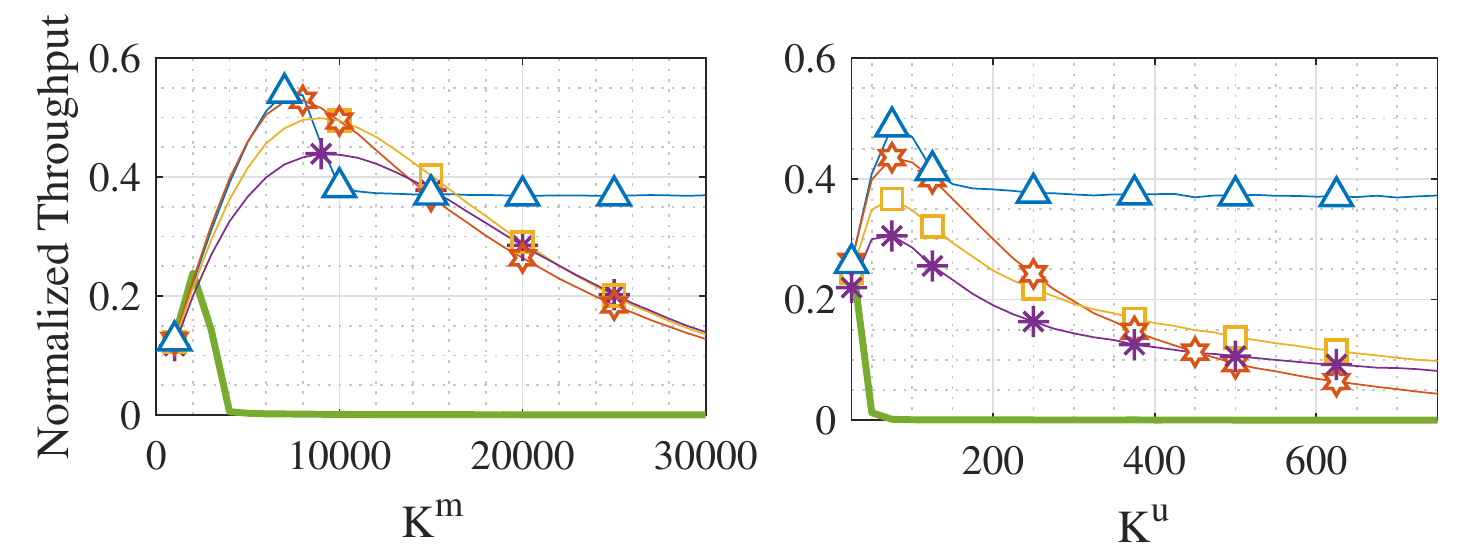}
\caption{{\bf Fixed} number of channels per user mode: {$L^u = 8$, $L^m=46$}.}
\label{fig:HGP_Fix}
\end{subfigure}
\hfill
\begin{subfigure}[!htbp]{0.49\textwidth}
\centering
\includegraphics[width=\textwidth]{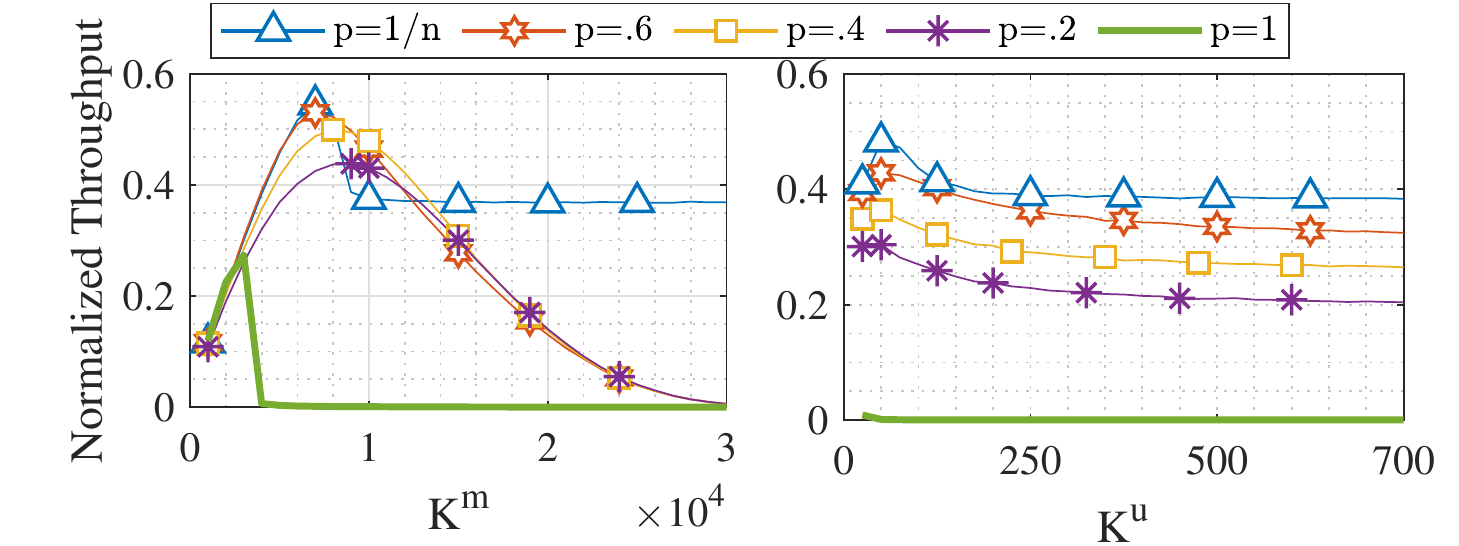}
\caption{{\bf Variable} number of channels per user mode: {$L^u \in \{4, 5, \dots, 33, 34\}$ as $K$ increases}} 
\label{fig:HGP_Var}
\end{subfigure}
\caption{\small Normalized Throughput $\eta$ for ACB-based contention resolution with $L = 54${, and $K^u = \frac{K^m}{{40}}$.}}
\label{fig:HGP_Preliminary}
\end{figure}

\vspace{-3mm}

Under perfect traffic prediction, successfully RS and applying the optimal ACB factor, Fig. \ref{fig:PTraffic_PAll} presents  $\eta$ metric and the  
average number of served users for $K^m$ up to $30,000$ users, {and $K^u = \frac{K^m}{400}$}, with BS having full knowledge of the oncoming traffic at every frame for both mMTC and URLLC modes. 
The $\eta_{\textsc{urlc}}$ is increased when the loading increases 
due to:  {\textit{i}})  less channel usage by the BS under low traffic, then the denominator in  eq. \eqref{eq:nThroughput} is  {reduced}, {\textit{ii}}) as $\Breve{K}^u$ increases, the number of {dedicated} channels {increases}, 
since the adopted access policy prioritizes URLLC mode. Additionally, given that the 
BS reserving resources independently of the mMTC traffic, and 
due to resources limitations, when $\Breve{K}^u$ is high, there is none or very few access opportunity to mMTC users, causing a high collision probability and abruptly decreasing the number of served users, as indicated in Fig. \ref{fig:HGP_ideal_servedUEs} for $\{K^m,K^u\} \geq \{10000,250\}$.

\begin{figure}[!htbp]
\vspace{-2mm}
\centering
\begin{subfigure}[!htbp]{0.48\textwidth}
\centering
 \includegraphics[width=\textwidth]{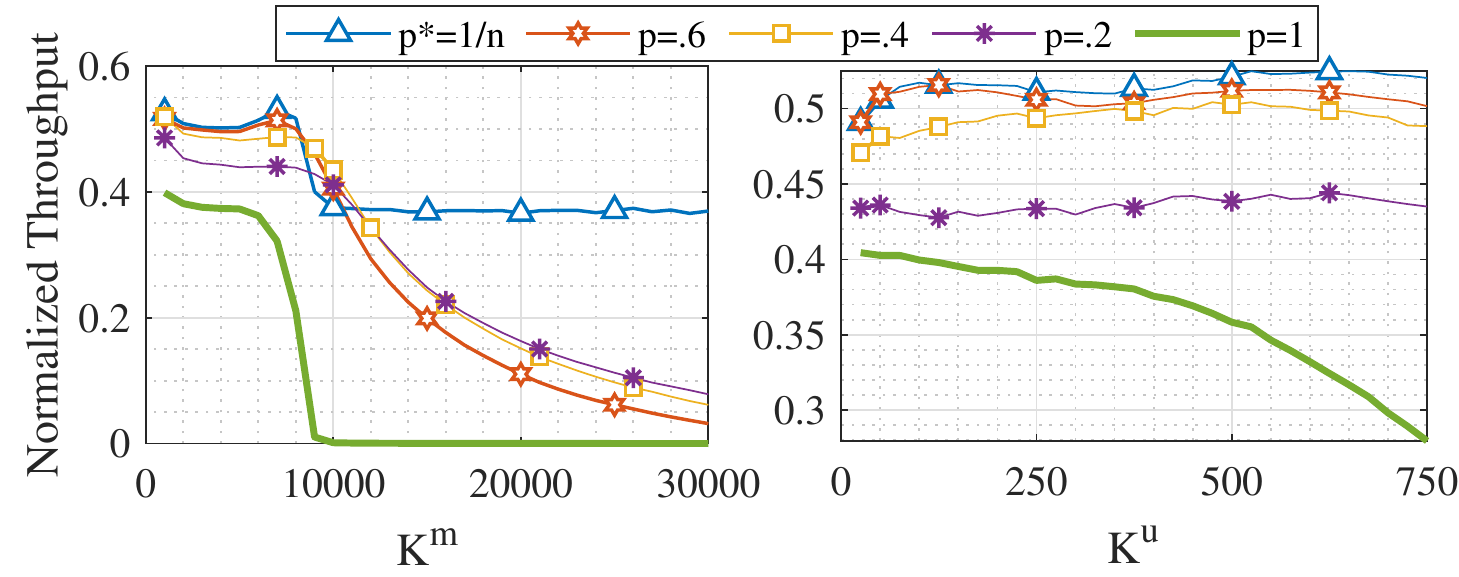}
 
 \vspace{-4mm}
\caption{Normalized throughput.}
\label{fig:HGP_ideal_throughput}
\end{subfigure}
\hfill
\begin{subfigure}[!htbp]{0.48\textwidth}
\centering
\includegraphics[width=\textwidth]{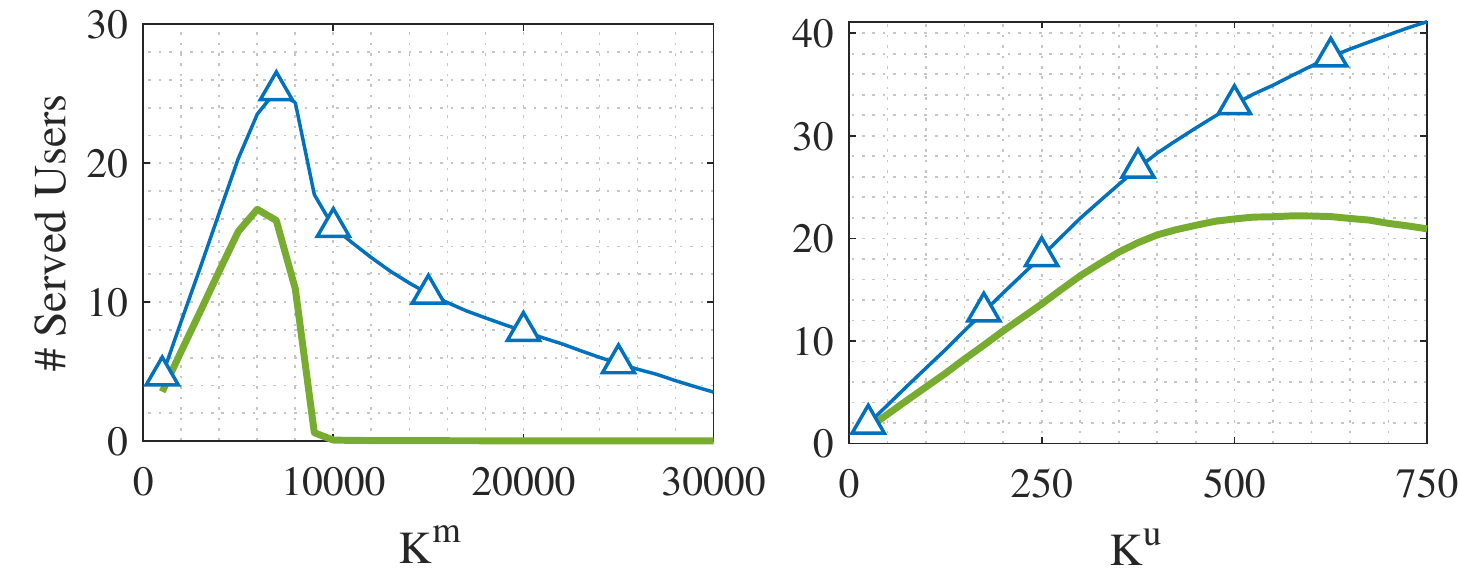}
\vspace{-3mm}

\caption{Mean number of served users.} \label{fig:HGP_ideal_servedUEs}
\end{subfigure}
\caption{\small Perfect RS and perfect traffic prediction for both URLLC and {crowded} mMTC {use modes. $\mathcal{F}=1200$ frames}; {$K^u = \frac{K^m}{400}$}}
\label{fig:PTraffic_PAll}
\end{figure}



%
%

\vspace{-1mm}
\bibliographystyle{IEEEtran}
\bibliography{ref.bib}

\end{document}